\documentclass[twocolumn,superscriptaddress,amsmath,aps]{revtex4}
\usepackage{graphicx,color}
\usepackage{bm}
\usepackage[hypertex]{hyperref}
\usepackage{float}

\newcommand{\be}{\begin{equation}}
\newcommand{\ee}{\end{equation}}
\newcommand{\bea}{\begin{eqnarray}}
\newcommand{\eea}{\end{eqnarray}}
\newcommand{\bsube}{\begin{subequations}}
\newcommand{\esube}{\end{subequations}}

\newcommand{\Eq}[1]{Eq.\,(\ref{#1})}
\newcommand{\Eqs}[1]{Eqs.\,(\ref{#1})}

\newcommand{\dg}{\dagger}
\newcommand{\la}{\langle}
\newcommand{\ra}{\rangle}

\newcommand{\nl}{\nonumber \\}

\newcommand{\beq}{\begin{equation}}
\newcommand{\eeq}{\end{equation}}
\newcommand{\beqn}{\begin{eqnarray}}
\newcommand{\eeqn}{\end{eqnarray}}
\newcommand{\bsub}{\begin{subequations}}
\newcommand{\esub}{\end{subequations}}

\begin{document}

\title{ Double-dot interferometer for quantum measurement
            of Majorana qubits \\ and stabilizers }
\author{Kai Zhou}
\affiliation{Department of Physics, Beijing Normal University, Beijing 100875, China}
\author{Cheng Zhang}
\affiliation{Department of Physics, Beijing Normal University, Beijing 100875, China}
\author{Lupei Qin}
\affiliation{Center for Joint Quantum Studies and Department of Physics,
School of Science, \\ Tianjin University, Tianjin 300072, China}
\author{Xin-Qi Li}
\email{xinqi.li@tju.edu.cn}
\affiliation{Center for Joint Quantum Studies and Department of Physics,
School of Science, \\ Tianjin University, Tianjin 300072, China}
\affiliation{Department of Physics, Beijing Normal University, Beijing 100875, China}

\date{\today}

\begin{abstract}
Motivated by the need of quantum measurement
of Majorana qubits and surface-code stabilizers,
we analyze the performance of a double-dot interferometer
under the influence of environment noise.
The double-dot setup design allows accounting for the full multiple
tunneling process between the dots through the Majorana island,
within a master equation approach.
In the co-tunneling regime, which results in a Majorana-mediated
effective coupling between the dots, the master equation approach
allows us to obtain analytic solutions for the measurement currents.
The measurement quality, characterized by figures of merit
such as the visibility of measurement signals,
is carried out in regard to the unusual decoherence effect
rather than `which-path' dephasing.
The results obtained in this work are expected to be useful
for future experiments of Majorana qubit and stabilizer measurements.
\end{abstract}
\maketitle

\section{INTRODUCTION}

In the past years, searching for Majorana zero modes (MZMs)
in topological superconductors has become an active field
in condensed matter physics \cite{Ali12,Bee13,Tew13,Nay15}.
Among the various efforts, the scheme based on semiconductor nanowires
has received particular attention because of the relative ease
of realization and control \cite{Kou12,Hei12,Xu12,Li13,Mar13,Kou18a}.
The key signatures of MZMs found in the experiments
are the zero-bias conductance peaks in tunneling spectra \cite{Kou12,Xu12,Kou18a}.

Beyond the problem of MZMs detection, it can be expected that the nanowire
based experiments will soon move to a more advanced level,
e.g., demonstration of non-Abelian statistics
and validation of a prototype topological qubit.  
The latter may include measuring the topological-qubit coherence times,
the residual MZM splittings, and quasiparticle poisoning rates.
All of these can be extracted from the quantum measurements of a Majorana qubit.

An even more advanced level of study is the fault-tolerant quantum computation,
in which the surface-code architectures are recognized
as very powerful platforms \cite{Fow12}.   
In particular, it has been recognized that
the Majorana surface code architecture
based on two-dimensional networks of nanowire MZMs
holds key advantages in both the hardware realization
and the actual operation of the code \cite{Egg16,Flen16}.
In this context, quantum measurements of the so-called stabilizers
in the surface code are the most important procedures,
which project the system to a well-defined code state
and are employed for error detection and correction
to realize fault tolerance in quantum operations.

In this work we focus on analyzing the measurement performance
of a double-dot interferometric device
shown in Fig.\ 1, which can be used for quantum measurement
of both Majorana qubits and stabilizers.
In this set-up, the Majorana island is coupled via two quantum dots
to the transport leads.
The interferometric device consists of two transmission paths,
say, a direct link between the two dots
and another one for electron flowing through the Majorana island.
The loop of interference enclosed by the two paths
is pierced by an external magnetic flux ($\phi$).
Via properly tuning the magnetic flux,
the different eigenvalues `$\pm 1$' of the Majorana qubit
or stabilizer will result in different measuring currents.

\begin{figure}  
   \centering
   \includegraphics[width=7.5cm]{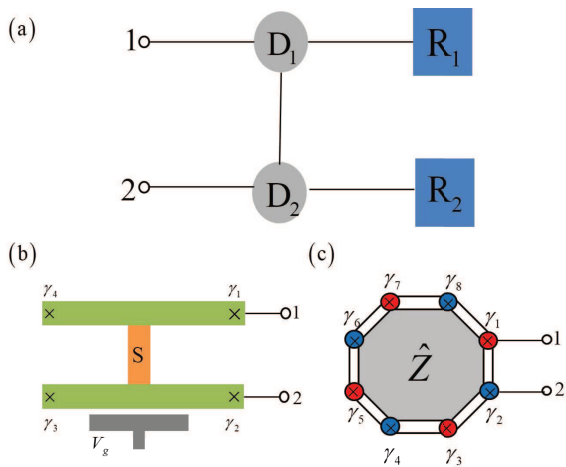}
\caption{
(a)
Schematic plot of the double-dot interferometer for
quantum measurements of Majorana qubits and stabilizers.
Inside the transport setup (with lead-reservoirs ${\rm R}_1$ and ${\rm R}_2$),
the double dots are tunnel-connected through two channels:
a direct link shown in the figure;
and another one through the Majorana island
in between the ports 1 and 2.
The two channels form an interference loop
through which a magnetic flux $\phi$ can be pierced.
(b)
A typical four Majorana-zero-modes qubit,
built with two topological-superconducting wires (green)
shunted by a conventional superconductor bridge (orange).
(c)
Example of a Majorana stabilizer designed in the Majorana surface codes.
The 8 Majoranas result in an effective code operator
$\hat{\cal Z}=\prod^8_{j=1}\gamma_j=\hat{z}_1\hat{z}_2\hat{z}_3\hat{z}_4$,
with the eigenvalues $Z=\pm 1$ to be determined by the quantum measurement.}
   \label{setup}
\end{figure}

For Majorana qubits,
in order to overcome limitations from parity conservation,
we consider a 4-MZMs qubit as shown in Fig.\ 1(b).
This Majorana box qubit (MBQ)
was recently described by Plugge {\it et al} \cite{Plu17}.
For Majorana stabilizers, we illustrate schematically in Fig.\ 1(c)
a concrete 8-MZMs plaquette, which was proposed
in Refs.\ \cite{Egg16,Flen16} in the nanowire-based Majorana surface codes.
Its measurement protocol was also proposed,
say, via a quantum interferometry by point contact measurements
of the linear conductance.

In this work, based on similar ideas we consider a specific set-up
where the Majorana island is coupled via quantum dots to the transport leads.
The main motivation of introducing
quantum dots in the interferometer set-up is as follows.
{\it (i)}
It allows us to account for the full multiple tunneling process
between the two dots through the Majorana island \cite{Qin19},
within a master equation approach.
If restricted in the lowest-order co-tunneling regime,
which results in a Majorana-island-mediated effective
coupling between the dots,
the master equation approach allows us to obtain analytic solutions
for the measurement currents, even in the presence of decoherence effects.
{\it (ii)}
The gate-controlled hybrid nanowires and quantum dots
are becoming very important platforms
for Majornana verification \cite{Cay15,Agu17,Vu18,Den18,Den16}
and quantum computation \cite{Flen16}.
For instance, in the recent experiments \cite{Den18,Den16},
the gate-controlled coupling structure of a nanowire and a quantum dot
was developed to probe the sub-gap Majorana states
via an improved technique of tunneling spectroscopy.
{\it (iii)}
Importantly, differing from most other quantum interferometers,
we will see that for our proposed set-up,
decoherence arisen from the dot-level fluctuations
{\it does not} seriously degrade the quality of measurement.
Under some parametric conditions, stronger decoherence of this type
can even enhance the measurement visibility,
quite outside a simple expectation.
{\it (iv)}
Finally, introducing the quantum dots in Fig.\ 1 may benefit
protecting the Majorana island from quasiparticle poisoning
(e.g. at finite temperature environments and finite bias voltage of transport),
compared to directly coupling the island to the continuum reservoirs/leads.

The paper is organized as follows.
In Sec.\ II we present the model description
and transport master equation approach
for the double-dot interferometer,
by considering to insert the Majorana qubit/stabilizer
into one of the interfering arms
(while remaining the derivation of the Majorana-island-induced
effective coupling in Appendixes A and B).
In Sec.\ III
we carry out analytic solution for the measurement currents
and compare it with the
numerical result from the full 12-state model of Majorana qubit.
We further analyze the decoherence effect in Sec.\ IV
and will pay particular attention to the unusual feature
owing to the dot-level fluctuations,
which would result in different behaviors
rather than the usual `which-path' dephasing.
We fianlly summarize the work in Sec V.

\section{Model and Methods}

\subsection{Set-up Description}

The double-dot interferometer we propose is schematically shown in Fig.\ 1,
where a Majorana island is inserted in one of the interference paths
while the other path is a direct tunnel-link between the dots.
The two quantum dots are further coupled to transport leads.
As usual, the interference loop enclosed by the two paths
is pierced by an external magnetic flux ($\phi$).
The phase-sensitive transport current between the leads
can reveal the state of the Majorana island.
Here, the Majorana island can be either a basic
qubit or a surface-code stabilizer.
As explained in the introduction, for the former, we consider a
4-MZMs qubit as shown in Fig.\ 1(b), which is built
by two topological superconductor nanowires
shunted by a conventional superconductor \cite{Plu17}.
For the stabilizer, we consider a 8-MZMs plaquette, which was proposed
in Refs.\ \cite{Egg16,Flen16} in the nanowire-based Majorana surface codes.

For the sake of completeness,
let us consider the Majornana qubit in a bit more detail.
The qubit consists of four MZMs, $\gamma_j=\gamma^{\dg}_j$, $j=1,\cdots,4$,
satisfying the fermion anticommutation relations
$\{\gamma_i,\gamma_j\}=2\delta_{i,j}$.
We assume long wires such that the MZMs have negligible overlap
and thus zero energy.
The Majorana fermions can be converted into the regular Dirac fermions through, e.g.,
$f^{\dg}_R=(\gamma_1-i\gamma_2)/2$ and $f^{\dg}_L=(\gamma_3-i\gamma_4)/2$.
The fermion parity of the 4-MZMs qubit reads
${\cal P}=\gamma_1\gamma_2\gamma_3\gamma_4=\pm 1$,
corresponding to an even/odd parity.
One can define the logical states of the qubit under,
for instance, the even parity with ${\cal P}=+1$ as
$|0\ra_L=|0_L,0_R\ra$, and $|1\ra_L=|1_L,1_R\ra$.
Here $0_{L/R}$ and $1_{L/R}$ denote, respectively, the empty and occupied states
of the left/right regular fermion described by $f^{\dg}_{L/R}$.
Accordingly, we can introduce the logical operator as
$\hat{z}=i\gamma_2\gamma_1$, together with
$\hat{x}=i\gamma_1\gamma_4$ and $\hat{y}=i\gamma_2\gamma_4$.

In order to prevent the qubit from quasiparticle poisoning,
an important design is building a small floating Majorana island
with large charging energy $E_C$.
The Coulomb interaction Hamiltonian is phenomenologically given by
$H_C=E_C(N-n_g)^2$, with $N$ the total charge on the floating island
and $n_g$ the gate charge which depends on the gate voltage
and gate capacitance. To minimize the charging energy,
the optimal choice is a symmetry point with an integer value of $n_g$.
More explicitly, we may denote the logical states of the qubit as
\bea
|0\ra_L &=& |0_L,0_R,N,n_0\ra \,, \nl
|1\ra_L &=& |1_L,1_R,N,n_0-1\ra \,.
\eea
Here $n_0$ denotes the number of Cooper pairs.
The total charge $N$ is conserved on the qubit Hamiltonian level
but varies during the transport measurement process.
That is, a single electron is allowed to {\it virtually}
enter or leave from the island.
For the $|0\ra_L$ sector, the relevant states include
\bea
&& |0_L,0_R,N,n_0\ra  \,, \nl
&& |0_L,1_R,N+1,n_0\ra  \,, \nl
&& |0_L,1_R,N-1,n_0-1\ra  \,.
\eea
Similarly, for the $|1\ra_L$ sector, the states include
\bea
&& |1_L,1_R,N,n_0-1\ra \,, \nl
&& |1_L,0_R,N+1,n_0\ra  \,, \nl
&& |1_L,0_R,N-1,n_0-1\ra   \,.
\eea
For the setup shown in Fig.\ 1,
the `central system' of transport
includes also the two quantum dots,
which are assumed with strong on-site Coulomb interaction
allowing each dot to be empty or singly occupied.
We thus have 4 dot basis states
$|0,0\ra$, $|0,1\ra$, $|1,0\ra$, and $|1,1\ra$.
Hence, the total set of basis states of the central system
are 12 product states formed by the 4 dot states
and the 3 states of the Majorana qubit island
for each of the  $|0\ra_L$ and $|1\ra_L$ sectors.

The tunnel coupling between the
dots and Majorana qubit are described by \cite{Qin19}
\bea\label{tunn-H-1}
H'_1= (-\lambda_0e^{i\phi}d^{\dg}_2d_1
+ \lambda_1d_1\gamma_1 + i\lambda_2d_2\gamma_2)+{\rm h.c.}  \,.
\eea
$d_{1,2}$ are the (annihilation) operators of the quantum dots.
The flux enclosed in the interference loop is accounted for
by the phase factor $e^{i\phi}$.
The asymmetric choice of the phases of the coupling amplitudes
between dots 1 and 2 and the respective Majoranas
is for a convenience to produce simple results
for real values of $\lambda_{0,1,2}$ \cite{Qin19}.

Finally, we describe the coupling between the dots and the transport leads by
$H'_2=\sum_k(t_1d^{\dg}_1c_{1,k}+t_2d^{\dg}_2c_{2,k}+{\rm h.c.})$.
The leads are assumed to be occupied with electrons
(with creation operators $c^{\dg}_{j,k}$)
depending on the bias voltages.
The continuum of the leads are characterized by the densities of states $\nu_j$,
resulting in coupling rates with the dots given by
$\Gamma_j=2\pi\nu_j|t_j|^2$.
In this work we consider $\Gamma_1=\Gamma_2=\Gamma$.

For the double-dot setup transport, one can conveniently simulate
the problem within a master equation approach, by
accounting for the full 12 or 24 states mentioned above,
which contains the full multiple tunneling process
between the dots through the Majorana island \cite{Qin19}.
However, as carefully exploited in Ref.\ \cite{Qin19},
under proper choice of parameters, the full 12-states problem
can be well approximated by the cotunneling process.
That is, under the condition $|\lambda_{1,2}|\ll E_C$,
the dominant cotunneling process will result in an
approximate low-energy effective Hamiltonian
for the coupling between the two dots as \cite{Plu17}
\begin{equation}\label{H-eff}
H'_{\rm eff}=\bigg(-\lambda_0e^{i\phi}+2\hat{z}\frac{\lambda_1\lambda_2^*}{E_C}\bigg)
d_2^\dagger d_1+{\rm h.c.}  \,.
\end{equation}
For a brief derivation of this result, see Appendix A of this article.

For the surface code stabilizer shown in Fig.\ 1(c),
applying the 4th-order perturbation theory (see Appendix B),
the effective low-energy code Hamiltonian of the 8-MZMs plaquette
reads \cite{Flen16}, $H_{\rm code}=-{\rm Re}(c) \hat{\cal Z}$,
with the stabilizer operator given by
$\hat{\cal Z}=\prod_{j=1,8}\gamma_j$
and the coefficient $c$ given by \Eq{eqHcode} in Appendix B.
The stabilizer operator has the same eigenvalues $\pm 1$
of the qubit operator $\hat{z}$.
However, corresponding to each of the eigenvalue $\pm 1$,
the individual states of the Majorana pairs,
i.e., the occupation of regular fermions, can be different.
The number of different states, after accounting also
for the single electron entering or leaving from the Majorana island,
are much larger than the qubit case.
This will make the full states simulation quite complicated.
Nevertheless, by applying as well the perturbation theory
(see Appendix B), one can obtain a similar low-energy tunneling
Hamiltonian as \Eq{H-eff},
needing only to replace the second term in the bracket of \Eq{H-eff}
by $\alpha(\xi+c^* \hat{\cal Z})d^{\dagger}_1d_2$.
For more details about $\alpha$ and $\xi$, see Appendix B.
Then, we see that the measurement problem of a Majorana qubit and a stabilizer
can be investigated by a unified treatment, i.e.,
based on the type of the effective tunneling Hamiltonian \Eq{H-eff}.

\subsection{Master Equation Approach}

For quantum transport,
the master equation approach is a very convenient tool.
From the perspective of quantum dissipation,
the transport leads can be regarded as a generalized fermionic environment,
and the `system of interest' is the central device,
e.g., the Majorana island plus the two quantum dots in our case.
For weak coupling between the central system and the transport leads,
under the Born approximation,
the so-called Born-Markov-Redfield master equation
for the reduced state of the central system reads \cite{Qin19,Li05,Luo07}
\begin{equation}
\dot\rho=-i\mathcal{L}_S\rho-
\frac{1}{2} \sum_{j=1,2}
\big\{[d_j^\dagger,D_j^{(-)}\rho-\rho D_j^{(+)}]+{\rm h.c.}\big\} \,.\label{eq1}
\end{equation}
Here, $\mathcal{L}_S$ is the Liouvillean superoperator
associated with the system Hamiltonian, $\mathcal{L}_S(\cdot\cdot\cdot)\equiv[H_S,(\cdot\cdot\cdot)]$.
In the latter dissipative terms, we have introduced:
\begin{equation}
D_j^{(\pm)}=\int_{-\infty}^{\infty}{\rm d}t C_j^{(\pm)}(t)
\big[e^{\mp i{\cal L}_S t}\,d_j \big]  \,.
\end{equation}
The reservoir correlation functions are defined from the respective
local-equilibrium thermal averages of the reservoir operators
\bea
C_j^{(+)}(t) &=& \sum_{k}|t_j|^2\big<c_{j,k}^\dagger(t)c_{j,k}(0)\big> \,,   \nl
C_j^{(-)}(t) &=& \sum_{k}|t_j|^2\big<c_{j,k}(t)c_{j,k}^\dagger(0)\big> \,.  \nonumber
\eea
More explicitly, under the wideband approximation for the leads, we have
$C_j^{(\pm)}(t)=|t_j|^2\sum_{k}e^{\pm i\varepsilon_kt} f_j^{(\pm)}(\varepsilon_k)$,
where $f_j^{(+)}(\varepsilon_k)=f_j(\varepsilon_k)$ is the Fermi function
of reservoir $j$, and $f_j^{(-)}(\varepsilon_k)=1-f_j(\varepsilon_k)$.
The Fermi functions depend on the respective electro-chemical potentials $\mu_j$.
Then in the eigenstate basis of $H_S$,
we can easily carry out the matrix elements of $C_j^{(\pm)}$ as
\bea
(D_j^{(\pm)})_{nm}
=\Gamma_j f_j^{(\pm)}(\omega_{mn})(d_j)_{nm} \,,
\eea
where $\omega_{mn}=E_m-E_n$ is the energy difference between the eigenstates of $H_S$.
Knowing the reduce state $\rho$ of the central device,
the transport current flowing into the drain reservoir
is given by \cite{Li05,Luo07}
\begin{equation}
I=\frac{1}{2}{\rm Tr}[(d_2^\dagger D_2^{(-)}
-D_2^{(+)}d_2^\dagger)\rho(t)+{\rm h.c.}]   \,.   \label{eq2}
\end{equation}
Here we may mention that, in addition to a convenient calculation
of the stationary current, the master equation approach
allows calculating the time-dependent currents
much more conveniently than the Green's function method
and the Landauer-B\"uttiker scattering approach.

\section{Measurement Currents}

To carry out explicit solutions, let us convert the operator form
of the master equation into a matrix-elements form,
by using the state basis of the system Hamiltonian $H_S$.
The most convenient choice of the state basis is
the number states of occupation of the double dots, i.e.,
$|1\rangle=|00\rangle$, $|2\rangle=|01\rangle$,
$|3\rangle=|10\rangle$, and $|4\rangle=|11\rangle$.
Here we only consider single occupation of each dot,
owing to the strong Coulomb blockade effect.
Also, the explicit occupation states of the Majorana island are not needed,
since we would like to adopt the description of the Majorana-mediated
effective coupling between the dots, given by \Eq{H-eff}.

Without loss of the essential physics and practical relevance,
let us consider also the zero temperature and large bias limits,
which result in $D_1^{(+)}=\Gamma_1d_1$,
$D_2^{(-)}=\Gamma_2d_2$, and $D_1^{(-)}=D_2^{(+)}=0$.
The large bias voltage means that the voltage window contains
all the four lowest eigen-energy levels,
but not including the high energy levels
resulting from the charging energy of the Majorana island.
More explanations are referred to Ref.\ \cite{Qin19}.
Based on these considerations, we find
\bea\label{Meq1}
\dot\rho_{11}=-\Gamma_1\rho_{11}+\Gamma_2\rho_{22} \,,\nonumber\\
\dot\rho_{22}=-i(\Omega\rho_{32}-\Omega^*\rho_{23})
-(\Gamma_1+\Gamma_2)\rho_{22}  \,,\nonumber\\
\dot\rho_{33}=-i(\Omega^*\rho_{23}-\Omega\rho_{32})
+\Gamma_1\rho_{11}+\Gamma_2\rho_{44},\nonumber\\
\dot\rho_{44}=\Gamma_1\rho_{22}-\Gamma_2\rho_{44} \,,\nonumber\\
\dot\rho_{23}=-i[-\delta\rho_{23}+
\Omega(\rho_{33}-\rho_{22})]-\frac{1}{2}(\Gamma_1+\Gamma_2)\rho_{23} \,.
\eea
Here we have introduced a notation for
the total coupling between the dots,
$\Omega= -\lambda_0e^{i\phi}+z\tilde{\lambda}_{12}$,
with $z=\pm 1$ and $\tilde{\lambda}_{12}=2\lambda_1\lambda_2^*/E_C$.

\begin{figure} 
\centering
\includegraphics[width=8.5cm]{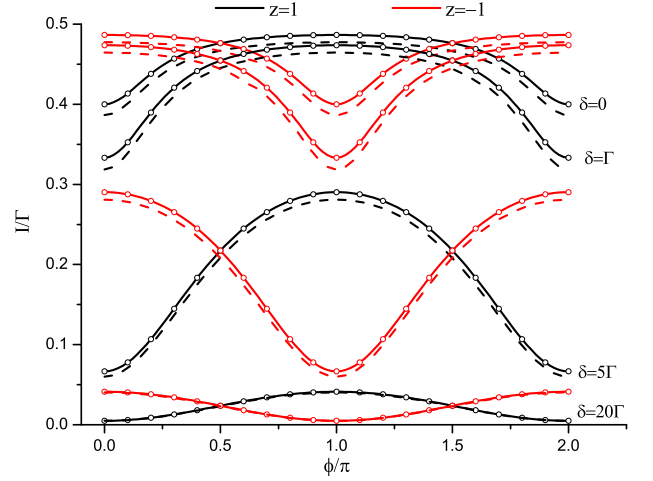}
\caption{
Measurement currents in steady state
for the distinct states $z=\pm 1$ of the Majorana qubit.
The interference effect is manifested through the dependence on
the enclosed magnetic flux.
The currents for $z=\pm 1$ differ by a phase shift of $\pi$.
The interference pattern is also sensitive to the
offset $\delta$ of the dot levels.
In the plot we compare results from several solving methods:
analytic solution (solid lines) and numerical result (circles)
from the effective coupling (4 states) model;
and numerical result (dashed lines) from the full 12-state model.
The parameters we used are
$\Gamma_1=\Gamma_2=\Gamma=0.01$, $E_C=100\Gamma$, $\lambda_0=\Gamma$,
and $\lambda_1=\lambda_2=10\Gamma$.   }
\end{figure}

In this work, we only consider the stationary currents of measurement,
which can be conveniently computed using \Eq{eq2}.
We find that only two matrix elements, say, $\rho_{22}$ and $\rho_{44}$,
are needed.
Their solutions read: $\rho_{44}=(\Gamma_1/\Gamma_2)\rho_{22}$, while
$\rho_{22}={4\Gamma_1\Gamma_2|\Omega|^2}\times {[(4|\Omega|^2+\Gamma_1\Gamma_2)(\Gamma_1+\Gamma_2)^2+4\Gamma_1\Gamma_2
\delta^2]^{-1}}$.
Then, from \Eq{eq2} we obtain
\bea\label{solution-1}
&& I(t\rightarrow\infty)
   =\Gamma_2(\rho_{22}+\rho_{44})   \nl
&& =\frac{4\Gamma_1\Gamma_2(\Gamma_1+\Gamma_2)|\Omega|^2}
   {(4|\Omega|^2+\Gamma_1\Gamma_2)(\Gamma_1+\Gamma_2)^2
   +4\Gamma_1\Gamma_2\delta^2}  \,.
\eea

In Fig.\ 2 we display the steady-state currents of measurement.
The quantum interference leads to a phase shift of $\pi$
between the currents for the Majorana qubit states $z=\pm 1$.
The effects of various parameters have been investigated
in detail in Ref.\ \cite{Qin19},
while here we only show the effect of the offset $\delta$ of the dot levels.
In Fig.\ 2 we compare results from several solving methods, say,
the analytic solution (solid lines) and numerical result (circles)
from the effective coupling (4 states) model,
and the numerical result (dashed lines) from the full 12-state model.
We find the results in satisfactory agreement.

Indeed, the 4-state low-energy Hamiltonian can reproduce
most results with sufficient accuracy \cite{Qin19}.
Differences only arise in some range of parameters,
associated with properties of the higher-energy states of the Majorana island.
Actually, in the full 12-state description, there are also the 8 high-energy states.
The eigenstates are superpositions of the low-energy and the high-energy
basis states, and even the low-energy eigenstates have a small,
but not exponentially suppressed contribution from the high-energy states and vice versa.
In Ref.\ \cite{Qin19} it was found that the current difference between
the full 12-state and effective 4-state treatments becomes pronounced
when $\lambda_0$ is large.
In this case, the total transmission is enhanced
but the interference effect is weakened
owing to the unbalanced transmission paths,
i.e., $\lambda_0 >> \tilde{\lambda}_{12}$.
However, after accounting for small finite temperature effect,
even this difference will be suppressed \cite{Qin19}.

\section{Decoherence Effects}

For the double-dot interferometer under consideration,
the study of decoherence effects can include such as
the energy fluctuation of the Majorana island,
the quasi-particle poisoning to the Majorana states,
and the level fluctuations of the double dots.
Owing to the strong Coulomb-blockade effect (large $E_C$),
we assume a strong suppression of the quasi-particle poisoning.
For the effect of the high-energy-level fluctuations of the Majorana island,
the full 12-states simulation (carried out in a separate work)
shows a slight reduction of the effective coupling between the dots
mediated by the Majorana island.
Therefore, in the present work, we would like to focus on
the decoherence effects of the double-dot level fluctuations.
Again, even in the presence of decoherence,
the low-energy effective coupling treatment
allows us to obtain analytic solutions.

\subsection{Naive Average}

Before analyzing the Lindblad-type decoherence effect,
we first calculate the average current
by averaging the dot-level fluctuations,
based on the analytic solution given by \Eq{solution-1}.
This manifests, in a rather transparent manner,
the physical picture of decoherence from the dot-level fluctuations.
We calculate
$\bar{I}=\frac{1}{2\Delta}
{\int_{\epsilon_1-\Delta}^{\epsilon_1+\Delta}I(\epsilon_1')
{\rm d}\epsilon_1'}$
and obtain
\begin{widetext}
\begin{equation}
\begin{split}
\bar{I}=&\frac{1}{2\Delta}
{\int_{\epsilon_1-\Delta}^{\epsilon_1+\Delta}\frac{4\Gamma_1\Gamma_2(\Gamma_1+\Gamma_2)|\Omega|^2}
{(4|\Omega|^2+\Gamma_1\Gamma_2)(\Gamma_1+\Gamma_2)^2+4\Gamma_1\Gamma_2
(\epsilon_1'-\epsilon_2)^2}{\rm d}\epsilon_1'}\\
=&\frac{\Gamma_1\Gamma_2|\Omega|^2}
{\Delta\sqrt{\Gamma_1\Gamma_2(4|\Omega|^2+\Gamma_1\Gamma_2)}}
\bigg\{\arctan\bigg[2\sqrt{\frac{\Gamma_1\Gamma_2}
{(4|\Omega|^2+\Gamma_1\Gamma_2)}}
\frac{\delta+\Delta}{\Gamma_1+\Gamma_2}\bigg]
-\arctan\bigg[2\sqrt{\frac{\Gamma_1\Gamma_2}
{(4|\Omega|^2+\Gamma_1\Gamma_2)}}
\frac{\delta-\Delta}{\Gamma_1+\Gamma_2}\bigg]\bigg\}.
\end{split}
\end{equation}
\end{widetext}

\begin{figure}
\centering
\includegraphics[width=8.5cm]{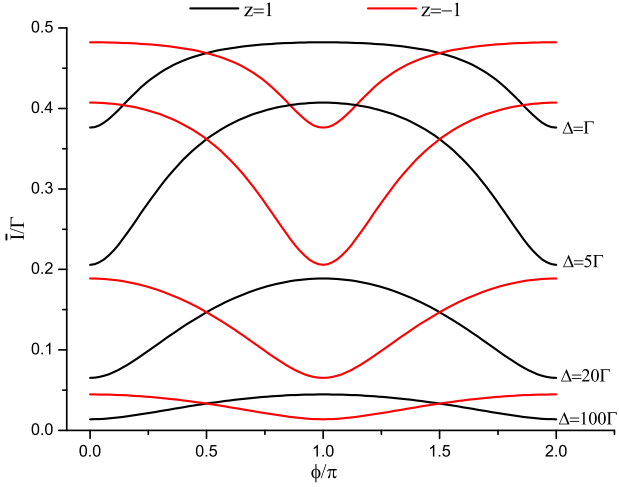}
\caption{ Average current over an amount $\Delta$
of the dot level ($\epsilon_1$) fluctuation.
The parameters are assumed as
$\Gamma=0.01$, $\epsilon_1=\epsilon_2=0$, $\lambda_0=\Gamma$,
and $\tilde{\lambda}_{12}=2\Gamma$.}
\end{figure}

In Fig.\ 3 we show the effect of the fluctuation amount ($\Delta$)
of the dot level ($\epsilon_1$) on the measurement current.
For any given magnetic flux ($\phi$),
we find that the average current decreases with $\Delta$.
The reason for this overall decreasing behavior is the decrease
of each individual current along with the increase of $\Delta$,
owing to stronger deviation from the resonant coupling under $\epsilon_1=\epsilon_2$.
However, the interference pattern does not vanish
or even is not reduced (e.g. see the case of $\Delta=5\Gamma$ in Fig.\ 3),
unlike the usual decoherence effect in the `which-path' interferometry set-up.
Here, the dot-level fluctuation does not correspond to the effect
of the `which-path' back-action.
For each pair of $(\epsilon'_1,\epsilon_2)$,
an interference pattern is indicated and most importantly,
the location of interference extrema of these patterns
does not shift with the difference of $\epsilon'_1$ and $\epsilon_2$.
Therefore, the in-phase summation (and average) of these patterns
does not cause {\it cancelation} which is actually
the origin of {\it dephasing} (decoherence) effect.
The naive treatment in this subsection has resulted in qualitatively similar result
as that in next subsection, using the Lindblad-type decoherence model
with constant rate ($\gamma$), e.g., compared with Fig.\ 4.
However, it seems not very clear how to {\it quantitatively} connect
the fluctuation amplitude ($\Delta$) analyzed here
with the decoherence rate $\gamma$ in the next subsection,
despite that for both treatments we can obtain analytic solutions.

\subsection{Lindblad-Type Decoherence}

Now we turn to the standard treatment of decoherence originated from
the energy level fluctuations of dot 1 with respect to dot 2,
under the influence of environment.
Under the wide-band limit (Born-Markovian approximation), we only need to modify
the transport master equation by adding a new term,
$\gamma D[\hat{s}]\rho$, with $\gamma$ the decoherence rate
and $\hat{s}=d^{\dagger}_1 d_1$ the decoherence operator.
In the same basis for \Eq{Meq1}, we have
\bea
\dot\rho_{23}&=&-i[-\delta\rho_{23}+\Omega(\rho_{33}-\rho_{22})]  \nl
&&  -\frac{1}{2}(\Gamma_1+\Gamma_2+\gamma)\rho_{23} \,.
\eea
The equations of other matrix elements remain the same as in \Eqs{Meq1}.

To obtain the steady-state current, we first carry out
the stationary solution of $\rho_{22}$ and $\rho_{44}$:
\bea
\rho_{22}&=& \frac{4|\Omega|^2\Gamma_1\Gamma_2(\Gamma_1+\Gamma_2+\gamma)}
{[B(\Gamma_1+\Gamma_2+\gamma)+ 4\delta^2\Gamma_1\Gamma_2]\,(\Gamma_1+\Gamma_2)} \nl
\rho_{44}&=& ({\Gamma_1}/{\Gamma_2})\,\rho_{22}  \,,
\eea
where we have introduced
$B=4|\Omega|^2(\Gamma_1+\Gamma_2)+\Gamma_1\Gamma_2(\Gamma_1+\Gamma_2+\gamma)$
to simplify the expression of $\rho_{22}$.
Based on this solution, the steady-state current can be obtained through
$I=\Gamma_2(\rho_{22}+\rho_{44})$, which yields
\bea\label{Is-2}
I(t\rightarrow\infty)
=\frac{4|\Omega|^2\Gamma_1\Gamma_2(\Gamma_1+\Gamma_2+\gamma)}
{B(\Gamma_1+\Gamma_2+\gamma)+4\Gamma_1\Gamma_2\delta^2} \,.
\eea

\begin{figure}
\centering
\includegraphics[width=8.5cm]{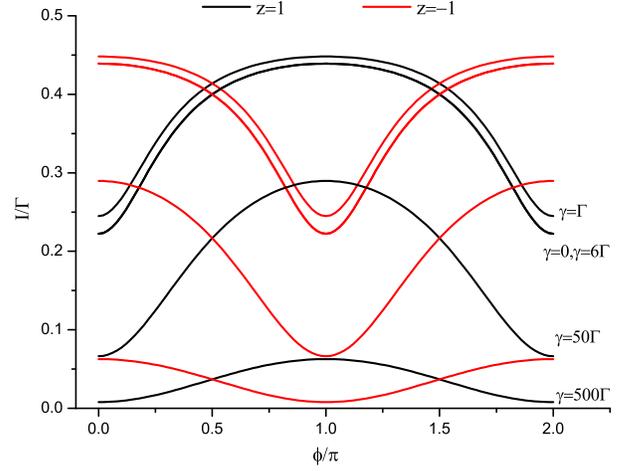}
\caption{
Decoherence effect on the measurement currents.
For the purpose discussed in the main text,
we set a nonzero offset of the dot levels ($\delta=2\Gamma$).
Other parameters are the same as in Fig.\ 2.}
\label{current-gamma}
\end{figure}

In Fig.\ 4 we plot the results based on \Eq{Is-2}, showing the decoherence effect
(i.e. the dot-level fluctuations) on the current.
The effect is qualitatively the same as the $\Delta$-effect
(the fluctuating amplitude of the dot level) shown in Fig.\ 3,
despite that $\gamma$ is determined
by the coupling strength to environment
and the density-of-states of the environment.
Indeed, $\gamma$ has the same effect of the level fluctuation amplitude.
In Fig.\ 4 we adopt $\delta=\epsilon_1-\epsilon_2=2\Gamma$, i.e.,
the two dots are not coupled in resonance.
We then find the current {\it increasing}
with the increase of $\gamma$ in the first stage, owing to
better matching of the dot levels caused by the stochastic fluctuations.
This behavior is somehow similar to the phenomenon of `dissipation-assisted tunneling'.
However, of course, the current would decrease with the further
increase of $\gamma$, because of the stronger deviation from resonance.

With the help of the analytic solution \Eq{Is-2},
we may determine some characteristic values of $\gamma$.
First, for a given $\delta$, we can find a $\gamma^*$,
which maximizes the current. From the following result
\bea
\frac{{\rm d}I}{{\rm d}\gamma}=\frac{4|\Omega|^2\Gamma_1^2\Gamma_2^2
[4\delta^2-(\Gamma_1+\Gamma_2+\gamma)^2]}
{\big\{B(\Gamma_1+\Gamma_2+\gamma)+4\Gamma_1\Gamma_2 \delta^2\big\}^2} \,,
\eea
we obtain $\gamma^*=2(|\delta|-\Gamma)$
by means of $\frac{{\rm d}I}{{\rm d}\gamma}|_{\gamma=\gamma^*}=0$.
Here we assumed $\Gamma_1=\Gamma_2=\Gamma$.
As a simple check, taking $\delta=2\Gamma$ as an example,
the current is enhanced with the increase of
$\gamma$ from zero to $2\Gamma$.
This is what we observed in Fig.\ 4.
The second characteristic value of our interest is $\gamma_0$,
given by $I(\gamma=0)=I(\gamma_0)$. We obtain
\begin{equation}
\gamma_0=\frac{4\delta^2-(\Gamma_1+\Gamma_2)^2}{\Gamma_1+\Gamma_2} \,.
\end{equation}
Again, for $\Gamma_1=\Gamma_2=\Gamma$ and $\delta=2\Gamma$,
we have $\gamma_0=6\Gamma$,
which is precisely the coincidence value found in Fig.\ 4.

\subsection{Interference Visibility}

As observed in Figs.\ 2 and 3,
the interference signal of the present double-dot interferometer
does not sensitively decrease with the decoherence strength between the two dots.
We are thus interested in a more meaningful quantity,
say, the {\it visibility} of the interference pattern, defined as
\begin{equation}
V=\frac{I_{{\rm max}}-I_{{\rm min}}}{I_{{\rm max}}+I_{{\rm min}}} \,.
\end{equation}
$I_{{\rm max}}$ and $I_{{\rm min}}$ are, respectively, the peak and valley
values of current oscillation with the magnetic flux $\phi$.
Obviously, the larger visibility means a better distinguishability
of the qubit states from the quantum measurement.
Based on the analytic solution we find
\bea\label{visib-1}
V=\frac{2\Gamma_1\Gamma_2 K |\lambda_0\tilde{\lambda}_{12}|}
{4(\Gamma_1+\Gamma_2)\tilde{\Gamma}
(\lambda_0^2-\tilde{\lambda}_{12}^2)^2
+\Gamma_1\Gamma_2 K (\lambda_0^2+\tilde{\lambda}_{12}^2)} \,.
\eea
Here we introduced $K=\tilde{\Gamma}^2+4\delta^2$,
and $\tilde{\Gamma}=\Gamma_1+\Gamma_2+\gamma$.

\begin{figure}
\centering
\includegraphics[width=8.5cm]{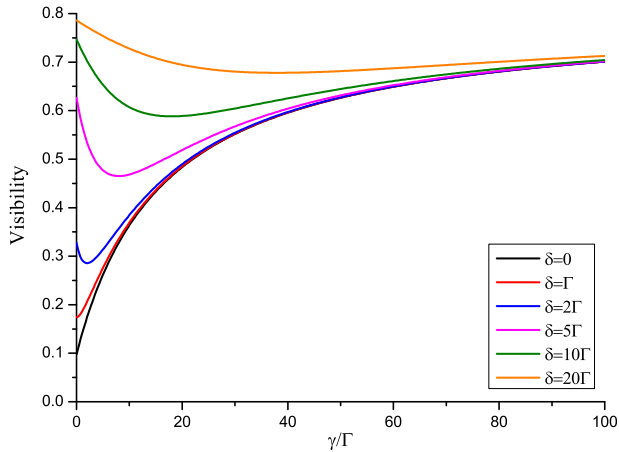}
\caption{
Decoherence effect on the visibility of the interference pattern,
e.g., shown in Fig.\ 4.
Results for different offset $\delta$ of the dot levels are displayed.
A unique feature is
the monotonically enhanced visibility (for small $\delta$)
along the decoherence rate $\gamma$,
which is unusual compared to the decoherence effect
in the usual `which-path' interferometry setups.
Parameters: $\Gamma=0.01$, $\lambda_0=\Gamma$, and $\tilde{\lambda}_{12}=2\Gamma$.}
\end{figure}

In Fig.\ 5 we show the {\it decoherence effect} on the visibility.
The most remarkable feature observed here is that the visibility
is enhanced (for small level-offset $\delta$)
along the decoherence rate $\gamma$, which is unusual
compared to the decoherence effect in other `which-path' setups.
The basic reason, as mentioned in Sec.\ IV A (associated with Fig.\ 3),
is that the dot-level fluctuation in our double-dot interferometer
does not result in the `which-path' back-action.
Moreover, for each `fluctuating configuration' of the dot-levels,
the quantum mechanically implied interference pattern
shares the same magnetic-flux ($\phi$) conditions
for constructive and destructive interferences.
Then, no decoherence-associated-cancelation occurs in the current.
However, the level fluctuations do affect (reduce)
the magnitude of current (for vanished or small $\delta$),
owing to driving the dots further away from the condition of resonant tunnel coupling.
Therefore, the interplay between the {\it persistently surviving}
quantum interference and the {\it level-fluctuation-induced}
off-resonance transmission results in the enhanced visibility
if we increase the decoherence rate $\gamma$, as seen in Fig.\ 5.
Similar explanation applies as well to the enhanced visibility
by increasing the level offset $\delta$,
for a given decoherence rate $\gamma$ as shown in Fig.\ 5.

More complicated situation occurs if we consider
to increase both $\delta$ and $\gamma$.
This will result in the turnover behavior observed in Fig.\ 5.
Based on the analytic result of \Eq{visib-1},
we find that the turnover point is given by
$\gamma=2|\delta|-\Gamma_1-\Gamma_2$.
Indeed, in Fig.\ 5 we find that for $\delta>\Gamma$
the visibility decreases first, then increases with $\gamma$.
But for $0\leq\delta\leq\Gamma$,
the visibility only increases with $\gamma$, monotonically.

\section{Summary}

We have analyzed the performance of a double-dot interferometer, in connection with
the quantum measurement of Majorana qubits and surface-code stabilizers.
The double-dot design has some advantages
such as separating the Majorana island from fermionic environment
(thus avoiding quasiparticle poisoning).
In the co-tunneling regime (through the Majorana island), the double-dot setup
allows an efficient low-energy effective description
for the Majorana-mediated coupling between the dots,
and a simple master equation approach allows us
to obtain analytic solutions for the transport currents
(even in the presence of environment noises).
A noticeable feature of the double-dot interferometer
is that the dot-level fluctuations do not cause
sensitive degradation of visibility of the measurement signals,
which is unusual regarding to the decoherence effect in other `which-path' setups.
The results analyzed in this work are expected to be useful
for future experiments of Majorana qubit and stabilizer measurements.

\vspace{1.0cm}
{\flushleft\it Acknowledgements.}---
This work was supported by the
National Key Research and Development Program of China
(No.\ 2017YFA0303304) and the NNSF of China (Nos.\ 11675016, 11974011 \& 61905174).

\appendix

\section{Effective Coupling Mediated by the Majorana Qubit}

In the Coulomb blockade regime (with large charging energy $E_C$),
the electron transmission through the Majorana island as shown in Fig.\ 1
involves only `virtual' occupation of the island states.
In this case, it is desirable to develop an effective description
for the Majorana-island-mediated coupling \cite{Flen16,Fu18}.
In this Appendix, we present a brief derivation for
the effective coupling Hamiltonian, \Eq{H-eff}.

Let us re-denote the tunnel-coupling Hamiltonian between the dots
(${\rm D}_1$ and ${\rm D}_2$)
and the two Majoranas ($\gamma_1$ and $\gamma_2$) as
\begin{equation}
H_t=\sum_{j=1,2} \big(\lambda_jd_j^\dagger e^{-i\varphi_j/2}\gamma_j
    +{\rm h.c.}\big)  \,,   \label{eqHt}
\end{equation}
where the coupling amplitude $\lambda_j$ and the dot-electron operators
are denoted as the same as in \Eq{tunn-H-1}.
Taking this coupling Hamiltonian as perturbation
and applying the 2nd-order perturbative expansion,
the transfer amplitude of an electron from ${\rm D}_1$ to ${\rm D}_2$
is give by $\la f| H_t {\cal G}_0  H_t  |i\ra$,
where the initial (final) state $|i\ra$ ($|f\ra$)
corresponds to the electron in the quantum dot ${\rm D}_1$ (${\rm D}_2$)
and the Majorana qubit island in the same ground state.
In the perturbative expansion,
${\cal G}_0$ is the free Green's function of the Majorana island.

For electron transmission through the Majorana qubit
from ${\rm D}_1$ to ${\rm D}_2$, there are two sequences:
{\it (i)} first ${\rm D}_1\to \gamma_1$ then $\gamma_2\to {\rm D}_2$;
{\it (ii)} first $\gamma_2\to {\rm D}_2$ then ${\rm D}_1\to \gamma_1$.
Moreover, both sequences
have the same intermediate-state energy offset of $E_C$,
which results in an energy denominator $-1/E_C$
in the perturbative amplitude (contributed by the free Green's function).
For each transmission sequence, after identifying the energy denominator,
one can reorganize the operator orders,
under the anticommutation rule of fermions.
For instance, one transmission sequence (the second one mentioned above)
is accounted for as
\begin{eqnarray}
&&(\lambda_1d_1\gamma_1)(-i\lambda_2^*\gamma_2d_2^\dagger)  \nonumber\\
&&=-\lambda_1\lambda_2^*(i\gamma_2\gamma_1)d_2^\dagger d_1   \nonumber\\
&&=-\lambda_1\lambda_2^* \hat{z}d_2^\dagger d_1  \,.
\end{eqnarray}
The transmission sequence {\it (i)} can be treated similarly
to give the same result.
Summing the two contributions,
we arrive at the effective coupling Hamiltonian of the two dots as
\begin{equation}
H_{{\rm eff}}'=\bigg(-\lambda_0e^{i\phi}+2\hat{z}\frac{\lambda_1\lambda_2^*}{E_C}\bigg)
d_2^\dagger d_1+{\rm h.c.}  \,,  \nonumber
\end{equation}
i.e., \Eq{H-eff} used in the main text.

\section{Effective Hamiltonians for Stabilizers}

\subsection{Effective Code Hamiltonian}

Following Refs.\ \cite{Egg16,Flen16}, a minimal Majorana stabilizer plaquette
involves eight Majoranas (MZMs), as schematically shown in Fig.\ 1(c).
The eight Majoranas are contributed from four Majorana islands
(not shown in Fig.\ 1(c)), while the neighboring Majorana islands
are linked by tunnel bridges, described by the tunnel-coupling Hamiltonian
\begin{equation}\label{eqHtun}
H'=-\frac{t_{ll'}}{2}\gamma_l\gamma_{l'}
e^{i(\varphi_l-\varphi_{l'})/2}+{\rm h.c.}  \,,
\end{equation}
where $t_{ll'}$ is the coupling amplitude between
the Majoranas $\gamma_l$ and $\gamma_{l'}$,
and $\varphi_l$ ($\varphi_{l'}$) the respective
superconducting phase of the Majorana island $l$ ($l'$).

The Majorana islands are prepared with strong charging energies,
described by the charging-energy Hamiltonian
$H'_C=\sum^4_{j=1}E_C(N_j-n_{g,j})^2$.
Following Ref.\ \cite{Egg16}, we assume an isotropic $E_C$ for all islands,
and the back-gate-controlled charges $n_{g,j}$
close to the integer charges $N_j$,
i.e., $n_{g,j}=N_j+\Delta n_{g,j}$ with $|\Delta n_{g,j}|\ll1$.
Therefore, the energy required to add (remove) a single charge
to (from) the $j$th island is $E_j^\pm=(1\mp2\Delta n_{g,j})E_C$.

Now we outline a derivation for the effective code Hamiltonian for the stabilizer,
which is an effective description of the tunnel-coupling-induced fluctuations
associated with the ground state $|G\ra$ of the four Majorana islands.
More specifically, let us consider the fluctuation process to transfer
an electron in anticlockwise direction to the initial island.
The transfer amplitude is give by
$\la G| H' {\cal G}_0  H' {\cal G}_0 H' {\cal G}_0 H'  |G\ra$,
where ${\cal G}_0$ is the Green's function associated with the `free' Hamiltonian
of the four Majorana islands.
This fluctuation process involves the tunnel-coupled Majorana pairs
$\gamma_7\gamma_8$, $\gamma_5\gamma_6$, $\gamma_3\gamma_4$, and $\gamma_1\gamma_2$.
For the convenience of description, we assume the Majorana pairs
$(\gamma_8,\gamma_1)$, $(\gamma_2,\gamma_3)$, $(\gamma_4,\gamma_5)$,
and $(\gamma_6,\gamma_7)$ belong to, respectively, the island 1, 2, 3, and 4.
Obviously, there are $4!=24$ tunnel-transfer sequences,
given by the different permutations of the four tunneling Hamiltonians
in the 4th-order perturbation expansion.
{\it (i)}
As an example, one sequence is
$\gamma_1\gamma_2\gamma_3\gamma_4\gamma_5\gamma_6\gamma_7\gamma_8$,
which corresponds to transferring an electron
first from island 1 to 4, next from 4 to 3, then from 3 to 2,
and finally back to island 1.
Associated with this sequence, the energy excitations of the intermediate states
are, respectively, $E_1^-+E_4^+$, $E_1^-+E_3^+$, and $E_1^-+E_2^+$.
Then, the energy denominator in the perturbative expansion
(contributed by the Green's functions) is given by
$1/[(E_1^-+E_2^+)(E_1^-+E_3^+)(E_1^-+E_4^+)]=1/(8E_C^3)$.
There are sixteen transfer sequences that share this energy denominator.
{\it (ii)}
Similar treatment allows us to account for the contribution from
a different type of transfer sequences, for instance,
$\gamma_1\gamma_2\gamma_5\gamma_6\gamma_3\gamma_4\gamma_7\gamma_8$.
The energy denominator of this type is $1/(16E_C^3)$
and there are 8 sequences.

Let us define the stabilizer operator as
\bea
\mathcal{\hat{Z}}=\prod_{j=1}^8\gamma_j \,,
\eea
which is the product of the eight Majorana operators comprising a minimal loop.
Based on the above result of fluctuation amplitude,
we arrive at an effective low-energy code Hamiltonian as \cite{Flen16}
\begin{equation}
H_{{\rm code}}=-{\rm Re}(c)\mathcal{\hat{Z}},\,\,
c=\frac{5}{16E_C^3}\prod t_{ll'}\,.   \label{eqHcode}
\end{equation}

\subsection{Effective Coupling Hamiltonian Mediated by a Majorana Stabilizer}

In this subsection we further derive the effective coupling
between the two quantum dots mediated by the Majorana stabilizer.
To be specific, let us consider an electron transfer from ${\rm D}_2$ to ${\rm D}_1$.
There are two types of paths: one is the shortcut
through the connection link between $\gamma_2$ and $\gamma_1$;
another one is along the stabilizer loop, i.e.,
$(\gamma_2,\gamma_3)\to(\gamma_4,\gamma_5)\to(\gamma_6,\gamma_7)\to(\gamma_8,\gamma_1)$.
The transfer amplitude for the former case
is given by the 3rd-order perturbative expansion
$\la f| H_t {\cal G}_0  H' {\cal G}_0 H_t |i\ra$,
while for the latter case it is given by the 5th-order expansion
$\la f| H_t {\cal G}_0 H' {\cal G}_0  H' {\cal G}_0 H' {\cal G}_0 H_t  |i\ra$.
For the sake of seeing more clearly the transfer sequence,
let us define the segment operators
$L_j = d_j^\dagger\gamma_je^{-i\varphi_j/2}$
and $A_{mn} = \gamma_m\gamma_ne^{i(\varphi_m-\varphi_n)/2}$,
extracted from the tunneling Hamiltonians (\ref{eqHt}) and (\ref{eqHtun}).

\renewcommand\tabcolsep{0.6cm}
\begin{table}[!h]
\caption{
All the 6 sequences and energy denominators
of an electron transfer through the shortcut
of the Majorana stabilizer from ${\rm D}_2$ to ${\rm D}_1$ as shown in
Fig.\ 1.  }
\begin{tabular}{ccc}
  \hline
  \hline
  No& sequences   & energy denominators \\
  \hline
  1 &$L_1A_{12}L_2^\dagger$ & $(E_1^+E_2^+)^{-1}$ \\
  2 &$A_{12}L_1L_2^\dagger$ & $-[(E_1^-+E_2^+)E_2^+]^{-1}$ \\
  3 &$L_2^\dagger A_{12}L_1$ & $(E_2^-E_1^-)^{-1}$ \\
  4 &$A_{12}L_2^\dagger L_1$ &$-[(E_1^-+E_2^+)E_1^-]^{-1}$ \\
  5 &$L_1L_2^\dagger A_{12}$& $-[E_1^+(E_1^++E_2^-)]^{-1}$ \\
  6 &$L_2^\dagger L_1 A_{12}$ & $-[E_2^-(E_1^++E_2^-)]^{-1}$ \\
  \hline
  \hline
\end{tabular}
\end{table}

{\it (i)}
For the shortcut case, the transfer process
involves each of the operators $L_1$, $L_2^\dagger$ and $A_{12}$ once,
and there are $3!=6$ sequences as shown
in Table \uppercase\expandafter{\romannumeral1}.
Summing up all contributions (i.e., the energy denominators), we obtain \cite{Flen16}
\begin{eqnarray}
\frac{(E_1^+-E_1^-)(E_2^+-E_2^-)}{E_1^+E_1^-E_2^+E_2^-}
=\frac{4}{E_C^2}\eta\,,   \\
\eta=\frac{\Delta n_{g,1}\Delta n_{g,2}}{(1-4\Delta n_{g,1}^2)(1-4\Delta n_{g,2}^2)}\,.
\end{eqnarray}
The asymmetry parameter $\eta$ depends on the
charge offset parameters $\Delta n_{g,l=1,2}$
and thus will be negligibly small with vanishing charge offsets.
Then, we obtain the effective tunneling Hamiltonian mediated
by the shortcut link between $\gamma_2$ and $\gamma_1$,
\begin{equation}
H_{12,{\rm eff}}'=-\frac{2\lambda_1\lambda_2^*t_{12}}{E_C^2}\eta
 \, d_1^\dagger d_2+{\rm h.c.}\,.
\end{equation}

{\it (ii)}
For the transfer path around the Majorana loop,
permutation of the 5 tunnel-coupling Hamiltonians
will generate $5!=120$ transfer sequences
(part of them are illustrated in Table II).
Notice that each sequence would result in the same product of operators,
$\mathcal{\hat{Z}}d_1^\dagger d_2$,
differing from each other only in amplitude
owing to the different energy denominators.
Summing up all the energy denominators, we find a very simple factor
\bea
\frac{16}{E_C^4}\Pi_{j=1}^41/{(1-4\Delta n_{g,j}^2)}  \,.   \nonumber
\eea
Ignoring the relative small quantity ($\Delta n_{g,j}^2$) in the denominators,
we obtain the effective coupling Hamiltonian mediated by the Majorana loop
\begin{equation}
H_{12,{\rm eff}}''=-\frac{2\lambda_1\lambda_2^*(t_{34}t_{56}t_{78})^*}{E_C^4}
\mathcal{\hat{Z}}d_1^\dagger d_2+{\rm h.c.}\,.
\end{equation}

Finally, summing the contributions of the two types of paths, we arrive at
\begin{equation}\label{H12-eff}
H_{12,{\rm eff}}=\alpha(\xi+c^*\mathcal{\hat{Z}})d_1^\dagger d_2+{\rm h.c.} \,,
\end{equation}
where $\alpha=-32\lambda_1\lambda_2^*/(5t_{12}^*E_C)$ \\
and $\xi=(5|t_{12}|^2/(16E_C))\eta$.
This is the result originally presented in Ref.\ \cite{Flen16}.
Summing this effective coupling (mediated by the Majorana stabilizer)
and the direct coupling between the two quantum dots,
we see that the measurement principle of the stabilizer operator $\hat{\cal Z}$
falls into the same category as described in the main text by \Eq{H-eff}.
In particular, if the charge offsets $\Delta n_{g,1}$ and $\Delta n_{g,2}$
are negligibly small, the $\xi$-term in \Eq{H12-eff} vanishes.

\renewcommand\tabcolsep{1.1cm}
\begin{table*}
\centering
\caption{Illustrative examples of the sequences and the associated
energy denominators, for an electron transfer from the quantum dot
${\rm D}_2$ to ${\rm D}_1$ through the main loop of Majorana stabilizer as shown in Fig.\ 1. }
\begin{tabular}{ccl}
  \hline
  \hline
  No & sequences  & energy denominators \\
  \hline
  1&$L_1A_{87}A_{65}A_{43}L_2^\dagger$& $(E_1^+E_4^+E_3^+E_2^+)^{-1}$\\
  2&$A_{87}L_1A_{65}A_{43}L_2^\dagger$& $[(E_1^-+E_4^+)E_4^+E_3^+E_2^+]^{-1}$\\
  3&$L_1A_{65}A_{87}A_{43}L_2^\dagger$& $[E_1^+(E_1^++E_4^-+E_3^+)E_3^+E_2^+]^{-1}$\\
  4&$A_{65}L_1A_{87}A_{43}L_2^\dagger$& $[(E_4^-+E_3^+)(E_1^++E_4^-+E_3^+)E_3^+E_2^+]^{-1}$\\
  5&$A_{87}A_{65}L_1A_{43}L_2^\dagger$& $[(E_4^++E_1^-)(E_1^-+E_3^+)E_3^+E_2^+]^{-1}$\\
  6&$A_{65}A_{87}L_1A_{43}L_2^\dagger$& $[(E_4^-+E_3^-)(E_1^-+E_3^+)E_3^+E_2^+]^{-1}$\\
  7&$L_1A_{87}A_{43}A_{65}L_2^\dagger$&$[E_1^+E_4^+(E_2^++E_3^-+E_4^+)E_2^+]^{-1}$\\
  8&$A_{87}L_1A_{43}A_{65}L_2^\dagger$&$[(E_1^-+E_4^+)E_4^+(E_2^++E_3^-+E_4^+)E_2^+]^{-1}$\\
  9&$L_1A_{43}A_{87}A_{65}L_2^\dagger$
  &$[E_1^+(E_1^++E_2^++E_3^-)(E_2^++E_3^-+E_4^+)E_2^+]^{-1}$\\
  10&$A_{43}L_1A_{87}A_{65}L_2^\dagger$
  &$[(E_2^++E_3^-)(E_1^++E_2^++E_3^-)(E_2^++E_3^-+E_4^+)E_2^+]^{-1}$\\
  11&$A_{87}A_{43}L_1A_{65}L_2^\dagger$
  &$[(E_1^-+E_4^+)(E_1^-+E_2^++E_3^-+E_4^+)(E_2^++E_3^-+E_4^+)E_2^+]^{-1}$\\
  12&$A_{43}A_{87}L_1A_{65}L_2^\dagger$
  &$[(E_2^++E_3^-)(E_1^-+E_2^++E_3^-+E_4^+)(E_2^++E_3^-+E_4^+)E_2^+]^{-1}$\\
  13&$L_1A_{65}A_{43}A_{87}L_2^\dagger$&$[E_1^+(E_1^++E_3^++E_4^-)(E_1^++E_2^++E_4^-)E_2^+]^{-1}$\\
  14&$A_{65}L_1A_{43}A_{87}L_2^\dagger$&$[(E_3^++E_4^-)(E_1^++E_3^++E_4^-)(E_1^++E_2^++E_4^-)E_2^+]^{-1}$\\
  15&$L_1A_{43}A_{65}A_{87}L_2^\dagger$&$[E_1^+(E_1^++E_2^++E_3^-)(E_1^++E_2^++E_4^-)E_2^+]^{-1}$\\
  16&$A_{43}L_1A_{65}A_{87}L_2^\dagger$&$[(E_2^++E_3^-)(E_1^++E_2^++E_3^-)(E_1^++E_2^++E_4^-)E_2^+]^{-1}$\\
  17&$A_{65}A_{43}L_1A_{87}L_2^\dagger$&$[(E_3^++E_4^-)(E_2^++E_4^-)(E_1^++E_2^++E_4^-)E_2^+]^{-1}$\\
  18&$A_{43}A_{65}L_1A_{87}L_2^\dagger$&$[(E_2^++E_3^-)(E_2^++E_4^-)(E_1^++E_2^++E_4^-)E_2^+]^{-1}$\\
  19&$A_{87}A_{65}A_{43}L_1L_2^\dagger$&$[(E_1^-+E_4^+)(E_1^-+E_3^+)(E_1^-+E_2^+)E_2^+]^{-1}$\\
  20&$A_{65}A_{87}A_{43}L_1L_2^\dagger$&$[(E_3^++E_4^-)(E_1^-+E_3^+)(E_1^-+E_2^+)E_2^+]^{-1}$\\
  21&$A_{87}A_{43}A_{65}L_1L_2^\dagger$&$[(E_1^-+E_4^+)(E_1^-+E_2^++E_3^-+E_4^+)(E_1^-+E_2^+)E_2^+]^{-1}$\\
  22&$A_{43}A_{87}A_{65}L_1L_2^\dagger$&$[(E_2^++E_3^-)(E_1^-+E_2^++E_3^-+E_4^+)(E_1^-+E_2^+)E_2^+]^{-1}$\\
  23&$A_{65}A_{43}A_{87}L_1L_2^\dagger$&$[(E_3^++E_4^-)(E_2^++E_4^-)(E_1^-+E_2^+)E_2^+]^{-1}$\\
  24&$A_{43}A_{65}A_{87}L_1L_2^\dagger$&$[(E_2^++E_3^-)(E_2^++E_4^-)(E_1^-+E_2^+)E_2^+]^{-1}$\\
  \hline
  \hline
\end{tabular}
\end{table*}

\end{document}